\documentclass[sigconf, nonacm]{acmart}

\usepackage[english]{babel}
\usepackage{blindtext}

\renewcommand\footnotetextcopyrightpermission[1]{} 
\setcopyright{none}

\acmDOI{}

\acmISBN{}

\usepackage{tikz}
\usepackage{amsmath} 

\usepackage[frozencache,cachedir=.]{minted} 
\usepackage{colortbl}

\usepackage{algorithm}
\usepackage[noend]{algpseudocode} %
\algrenewcommand\algorithmicindent{0.5em}
\usepackage{etoolbox}
\usepackage{array}
\usepackage{threeparttable}
\usepackage{xspace}

\usepackage{booktabs}

\newcolumntype{x}[1]{>{\centering\let\newline\\\arraybackslash}p{#1}}

\makeatletter
\newcommand{\ssymbol}[1]{^{\@fnsymbol{#1}}}
\makeatother

\hypersetup{
    colorlinks,
    linkcolor={red!50!black},
    citecolor={blue!50!black},
    urlcolor={blue!80!black}
}

\newcommand{\name}{NeVerMore}
\newcommand{\namefull}{\name: Exploiting RDMA Mistakes in NVMe-oF Storage Applications}

\newcommand{\ta}{TLU\xspace} 
\newcommand{\tb}{TRA\xspace} 

\definecolor{ben}{RGB}{234, 127, 43} 
\definecolor{konst}{RGB}{43, 131, 186} 
\definecolor{adrian}{RGB}{43, 131, 186}
\definecolor{htor}{RGB}{43, 131, 186}

\usepackage{pifont}
\newcommand{\cmark}{\ding{51}}%
\newcommand{\xmark}{\ding{55}}%

\usepackage{url}

\usepackage{enumitem}

\begin{document}
\date{}

\title{\namefull} 
\author{  Konstantin Taranov,
Benjamin Rothenberger,
Daniele De Sensi, 
Adrian Perrig,
Torsten Hoefler}

\affiliation{ktaranov@inf.ethz.ch, 
rothenbb@inf.ethz.ch,
ddesensi@inf.ethz.ch,
aperrig@inf.ethz.ch,
htor@inf.ethz.ch}

\affiliation{%
\institution{ETH Zurich, Switzerland}
  }
 
\renewcommand{\shortauthors}{Taranov, et al.}


\begin{abstract}
This paper presents a security analysis of the InfiniBand architecture, a prevalent RDMA standard, and NVMe-over-Fabrics (NVMe-oF), a prominent protocol for industrial disaggregated storage that exploits RDMA protocols to achieve low-latency and high-bandwidth access to remote solid-state devices.
Our work, \name, discovers new vulnerabilities in RDMA protocols that unveils several attack vectors on RDMA-enabled applications and the NVMe-oF protocol,
showing that the current security mechanisms of the NVMe-oF protocol do not address the security vulnerabilities posed by the use of RDMA.
In particular, we show how an unprivileged user can inject packets into any RDMA connection created on a local network controller, bypassing security mechanisms of the operating system and its kernel, and how the injection can be used to acquire unauthorized block access to NVMe-oF devices. 
Overall, we implement four attacks on RDMA protocols and seven attacks on the NVMe-oF protocol and verify them on the two most popular implementations of NVMe-oF: SPDK and the Linux kernel.
To mitigate the discovered attacks we propose multiple mechanisms that can be implemented by RDMA and NVMe-oF providers.
\end{abstract}

\maketitle

\section{Introduction}
Resource disaggregation is becoming an important tool in data center design, splitting existing monolithic servers into a number of consolidated single-resource pools that communicate over high-speed interconnects.
This approach improves the hardware resource utilization and deployment flexibility as both the compute and storage nodes can use different types of server hardware and can be dimensioned independently. 
%
Despite these merits, disaggregation opens up new attack vectors, as it is often implemented over low-latency, high-bandwidth but untrusted  networks. 

The NVMe over Fabrics (NVMe-oF) protocol~\cite{nvmeof} is a leading protocol for  storage disaggregation, and it is offered and maintained by numerous storage and network vendors (Intel~\cite{spdk}, Xilinx~\cite{xilinx-nvmeof}, Mellanox~\cite{bluefiled}, Broadcom~\cite{ps225}, and Pure Storage~\cite{purestorage}). NVMe-oF combines two recent high-performance techniques: NVM Express (NVMe)~\cite{nvme} and Remote Direct Memory Access (RDMA).  NVMe-oF adopts RDMA connections to send NVMe requests, which are usually sent over PCIe to a local solid state drive (SSD), over a networking fabric with ultra-low latency of a few microseconds.
Even though RDMA networks enable low-latency and high-bandwidth, they have been shown to suffer from security weaknesses~\cite{srdma,redmark,Simpson,tsai2019pythia}. Key reasons are the lack of secure channels and the exposure of memory access to remote parties.  Despite these risks, the security implications and dangers of deploying NVMe-oF remain largely unstudied.

In this work, we introduce \emph{a series of attack tools that can be employed to attack any RDMA-enabled system} (see Section~\ref{sec:rdma}). We show that any system that tries to make use of RDMA opens an attack surface allowing local users to bypass the security mechanisms of the operating system and its kernel. Importantly, we show that any unprivileged user can inject packets into RDMA connections created on a local network controller, even if they are created in kernel space. Hence, any system that uses RDMA from the kernel space opens an attack surface allowing local users to manipulate RDMA-enabled kernel modules, such as the NVMe-oF block device. 
In addition, we show how an adversary can conduct denial-of-service attacks by breaking, preventing, and slowing down RDMA connections through vulnerabilities in the RDMA connection manager, RDMA resource sharing, and RDMA congestion mechanisms. 

Given the discovered RDMA vulnerabilities, we then analyze the security mechanisms of the NVMe-oF protocol as well as its implementations in the Storage Performance Development Kit (SPDK)~\cite{spdk} and the Linux kernel~\cite{nvme-linux} (see Section~\ref{sec:nvmeof}). Our analysis, \name, covers a recently proposed security extension~\cite{nvme,nvmeof} to NVMe-oF, which includes in-band authentication and secure channels using IPsec. 
We show multiple vulnerabilities and flaws in the design and implementation of the NVMe-oF protocol that are related to the use of RDMA networking.
In particular, we show how \emph{unprivileged users can acquire block access to NVMe-oF devices} and manipulate stored and loaded data.

%
Consequently, although the storage industry is spending much effort on implementing these security mechanisms for NVMe-oF, vulnerabilities in the underlying implementation and design of the InfiniBand architecture for RDMA networking   can impair these efforts.
Thus, we believe that the security issues inherited by using RDMA should be stressed to further influence the development of NVMe-oF security mechanisms.

\begin{figure}[t] 
\centering
\includegraphics[width=1\linewidth]{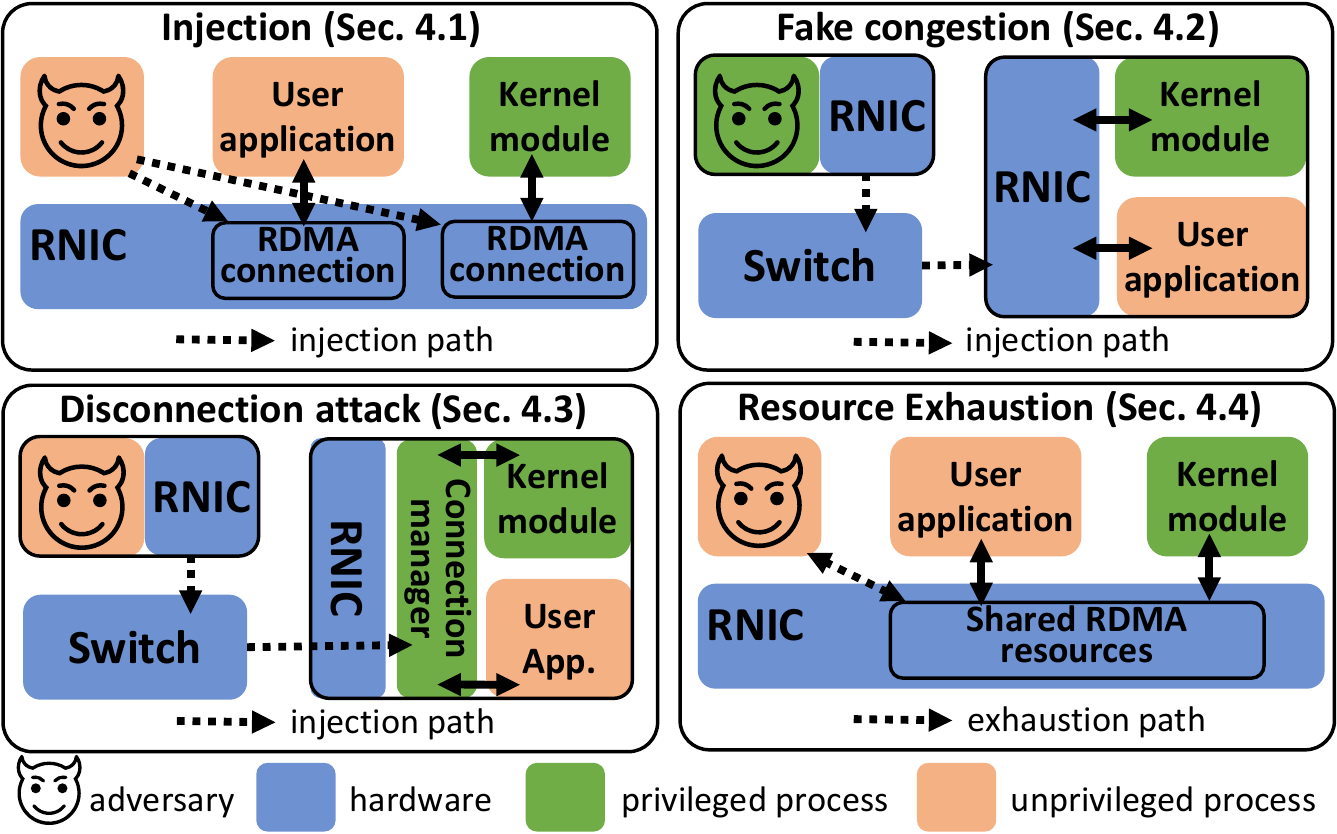}
\caption{Overview of proposed RDMA attacks.}
\label{fig:attacks}
\end{figure}

In summary, we propose four classes of attacks on RDMA protocols (see Figure~\ref{fig:attacks}) that enable seven different attack vectors on the NVMe-oF protocol, 
allowing an unprivileged adversary to manipulate the memory state of a remote NVMe device as well as any NVMe-oF client. In addition, we show how an adversary can severely destabilize the performance and the usability of the NVMe-oF protocol by exploiting vulnerabilities in the RDMA connection and resource management. 
To show the feasibility of the discovered attacks in practice, our attack framework, \name, was tested on the two most popular implementations of NVMe-oF. Finally, we discuss potential mitigation techniques for each of the discovered vulnerabilities, including a proposal for application-level source and data authentication for NVMe-oF requests (see Section~\ref{sec:mitig}).

\subsection{Overview of implemented attack classes}
\textbf{Injection}. An unprivileged user can inject packets into any RDMA connection created on a local network controller, including connections created by privileged kernel modules (see Section~\ref{sec:inject}). Therefore, any kernel application that makes use of RDMA opens an attack surface allowing the attacker to manipulate the kernel-level applications from user space by injecting RDMA requests into their connections. For NVMe-oF, we show in Section~\ref{sec:nvmemem} how the adversary can bypass security mechanisms of operating and file systems to directly manipulate NVMe disks at the block level without administrative privileges.

\textbf{Fake congestion}. A privileged user can forge congestion notification packets of RDMA protocols (see Section~\ref{sec:slowdown}), forcing remote network controllers to slow down. Therefore, the attacker can disrupt the normal operations of any reachable RDMA-enabled application. For NVMe-oF, the attacker can significantly degrade the performance of accesses to remote NVMe disks.  

\textbf{Disconnection attack}.  An unprivileged user can forge packets of RDMA connection manager (see Section~\ref{sec:rdmacm}), allowing it to disconnect any RDMA connection in the network, including connections created by local and remote kernel modules. 
Therefore, any user can disrupt the normal operations of any RDMA-enabled application. For NVMe-oF, the attacker can temporally disconnect network-attached NVMe disks, preventing the operating system from accessing them. 

\textbf{Resource exhaustion}. An unprivileged user can block local RDMA resources (see Section~\ref{sec:exhaust}), preventing local applications from opening RDMA connections.
Therefore, any local user can disrupt the normal operations of any local RDMA-enabled application. For NVMe-oF, the attacker can disconnect network-attached NVMe disks using the previous attack and then prohibit them from being reconnected, preventing the operating system from accessing storage for an extended period.

\section{Background on NVMe over RDMA}\label{sec:background}
\textbf{RDMA.} The NVMe-oF protocol uses RDMA network protocols specified in the InfiniBand architecture~\cite{infiniband2000infiniband} that includes native InfiniBand, RoCEv1, and RoCEv2 protocols.
Regardless of the underlying RDMA protocol, developers make use of RDMA communication through the RDMA verbs user space library~\cite{rdmacore}. 
Each reliable RDMA connection consists of two endpoint handlers called queue pairs, that allow applications to issue RDMA communication requests. 
Users submit asynchronous communication requests directly to the RDMA-capable network controller (RNIC) through its queue pair handler, bypassing the operating system and reducing CPU overhead. The RNIC performs all data accesses using an integrated DMA module that can directly write into and read from local memory.
Once the RNIC finishes the execution of a communication request, it generates a completion event that is written to a user space completion queue indicating completion of the request.

\textbf{NVMe.}
The NVMe protocol~\cite{nvme} is a storage protocol designed to take advantage of fast PCIe interfaces to send requests directly to high-performance storage media.  The NVMe protocol is similar to an RDMA protocol with the communicating endpoints being the host CPU and the storage device. Applications make use of NVMe by directly posting asynchronous work requests to a storage device that uses specialized DMA hardware to access memory.
As a result, NVMe provides microsecond-scale access latencies and millions of I/O operations per second, significantly outperforming legacy storage protocols such as SAS~\cite{seagate-sas} and SATA~\cite{sata}.

\textbf{NVMe-oF.}
The NVMe-oF protocol~\cite{nvmeof} defines a common architecture that supports NVMe block storage over a network. As a network protocol, NVME-oF leverages RDMA that offloads data movement to RDMA-capable network cards and, therefore, reduces the processing overheads involved in handling remote I/O requests. 
The NVMe-oF protocol encapsulates the NVMe commands and responses into a fabric-neutral capsule and passes it to the RDMA transport. A capsule represents the NVMe unit transferred from an NVMe-oF client to a remote NVMe-oF target that has a locally attached NVMe device (i.e., over PCIe). NVMe-oF read and write commands are discussed in detail in Section~\ref{sec:nvmeof}. 

\textbf{RDMA connection manager.}
The NVMe-oF client and the remote NVMe-oF target need to establish an RDMA connection to ensure reliable delivery of NVMe capsules. Connections are established using the RDMA connection manager, a user space library and a kernel module, designed to make an RDMA connection establishment and management similar to TCP sockets.
Otherwise, the RDMA application would have to implement an exchange of connection parameters over TCP connections and manually create RDMA endpoints using the RDMA verbs library, which is prone to errors and complicates the maintenance of RDMA connections. 

\begin{figure}[t] 
\centering
\includegraphics[width=0.95\linewidth]{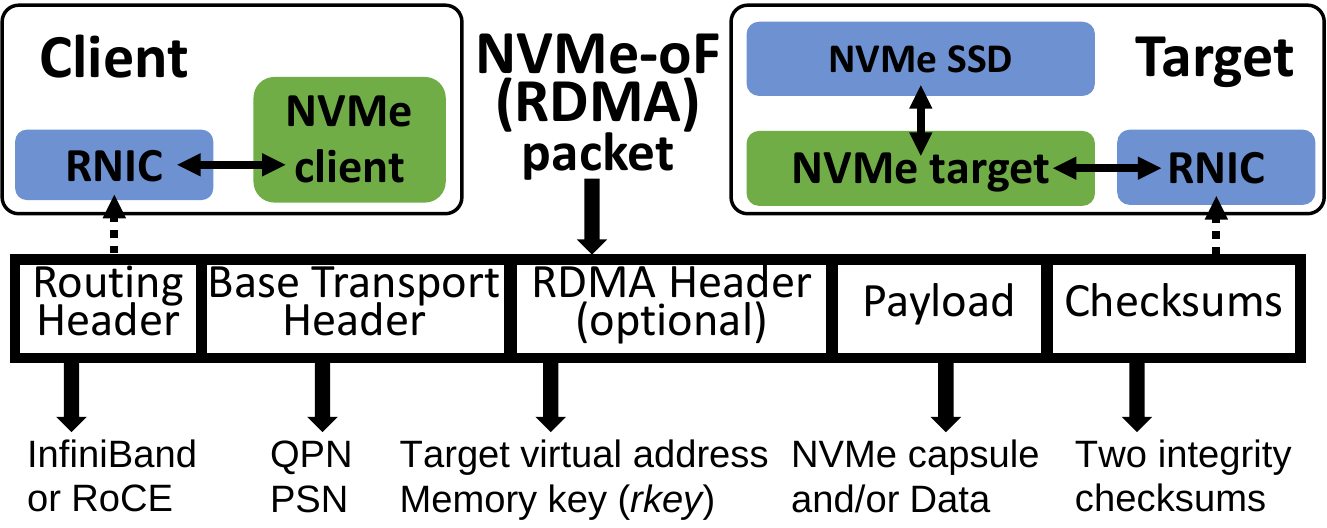}
\caption{NVMe-oF client-target model, and a packet format of NVMe-oF packet.}
\label{fig:overview}
\end{figure}

\textbf{End-to-end overview.}
Figure~\ref{fig:overview} illustrates a deployment example of NVMe-oF.
A storage node (target) runs an NVMe-oF target application that is privileged to access local NVMe devices via the NVMe protocol and listens for RDMA connections from other cluster nodes. A compute node (client) runs an NVMe-oF client application that connects to the target using the RDMA connection manager. The client can either be a kernel module that locally mounts a remote disk or a privileged application that is allowed to connect to a remote disk. Currently, NVMe-oF targets can use IP filters or an in-band authentication protocol~\cite{nvmeof,nvme-linux-patch}, offering bi-directional challenge–response authentication, to prevent unauthorised connections.

The NVMe-oF protocol uses RDMA \emph{send} for NVMe-oF requests and responses to notify a target node through completion events about incoming messages, and one-sided RDMA \emph{write} and \emph{read} requests for data communication to ``silently'' access data without generating completions events on the target. Here we give an overview of how RNICs process packets. Later we discuss each NVMe-oF request in detail in Section~\ref{sec:nvmeof}. When a client sends an NVMe request, it gets  encapsulated into an RDMA payload and sent to the remote target. Each RDMA packet consists of a routing header (containing network port identifiers), a transport header (containing information relevant to the endpoint such as connection identifier (QPN) and packet sequence number (PSN)), the packet payload, and two integrity checksums. 

The RNIC processes the RDMA packet in the following order. First, it checks the checksums and whether the targeted connection endpoint exists. For that, the RNIC checks whether a local table of open connections contains the concatenation of the routing information and the target connection identifier (QPN). Note that the RNIC does not only match QPN but also the routing information (e.g., the IP address of the sender for RoCEv2).
If the connection exists, the RNIC compares the packet sequence number (PSN) of the packet and the local packet counter. If they match, the RNIC copies the payload containing the capsule to the memory location specified by the target application in a pre-posted RDMA receive request. The application then processes the capsule and sends the response to the client.

Data communication in the NVMe-oF protocol is conducted through RDMA write and read requests, that additionally contain an RDMA header containing a memory key (\textit{rkey}) and a target virtual address.
To restrict unauthorized memory access, RDMA-enabled applications need to explicitly register memory for RDMA accesses to generate its memory key, which should be included in the request by the initiator.
To process one-sided RDMA packets, the RNIC verifies the RDMA header by checking the permissions of the targeted connection identifier (QPN) to access specified memory given the provided memory key. After these checks, the RNIC accesses the requested virtual address using its DMA engine. 

\section{Threat Model}

We consider two threat models which we denote as \ta and \tb. 
In both models, we consider a victim connection that connects two endpoints located on separate machines. The connection can be any RDMA connection based on the InfiniBand architecture such as RoCEv1, RoCEv2, and InfiniBand. In the case of NVMe-oF storage disaggregation, the connection is established between an NVMe-oF client and an NVMe-oF target. Furthermore, attacks that target the RDMA connection manager, assume that the connection is established using the RDMA connection manager. For all other attacks, the connection can either be established using the RDMA connection manager or using the native interface through the RDMA verbs library. Note that NVMe-oF always uses the RDMA  connection manager to establish connections.

\textbf{Threat model Local User (\ta).} We consider an adversary that is on one of the endpoints of the victim connection (i.e., it is co-located with either the NVMe-oF target or client). The attacker is an unprivileged user and is assumed to have obtained access to the machines using legitimate means. We assume that the attacker shares the same physical RNIC as the NVMe-oF entity and both can use it for communication. We assume that the attacker and the NVMe-oF entity are not separated through RNIC virtualization. The TLU model is prevalent in private clusters that use RDMA and NVMe-oF to accelerate their workloads. 

The link between the NVMe-oF entities could be secured using IPsec over RoCE~\cite{ipsecroce-slides}, a protocol that encapsulates RoCEv2 packets into IPsec packets, and a corresponding IPsec policy that was configured by a local administrator.
Neither the NVMe-oF entities nor the attacker has access to the cryptographic keys used for the IPsec security policy but can use the secure link to communicate to the remote entity. Consequently, all attacks implemented under
the assumption of the threat model TLU can be also performed when the link between the machines is not secured using IPsec.

\textbf{Threat model Remote Administrator (\tb).}
We assume that the attacker is located on a different machine than the endpoints of the NVMe-oF connection. The attacker has administrative privileges on its machine which allows it to fabricate and inject messages into the network. These privileges allow the attacker to change the configuration of the network interface (e.g., its IP address).  This model was proposed in ReDMArk~\cite{redmark} for packet injection, where it was called the T2 model. 

We assume that the IPsec over RoCE is not enabled for the link between the NVMe-oF entities. This allows the attacker to send forged packets to the endpoints of the victim connection. Consequently, all the attacks that are implemented within this model can be mitigated by enabling the IPsec channel over the path used by our two victim endpoints.

\textbf{Adversary constraints.}
We assume  the NVMe-oF target only accepts connections from benign NVMe-oF clients. Thus,  the adversary cannot directly establish a connection with the NVMe-oF target. Existing NVMe target applications can use IP filters for that. In addition, the security extension to NVMe-oF that is currently under development is assumed to prevent all unauthorized connections using the challenge-handshake authentication protocol~\cite{chap}.

We assume that the adversary is further constrained by not being able to eavesdrop on existing connections. For example, instead of sniffing RDMA connection parameters from existing connections (e.g., connection identifiers or packet sequence numbers), the attacker is required to guess these parameters in order to successfully impersonate one of the NVMe-oF endpoints. Guessing these parameters is facilitated using the attacks discovered in ReDMArk~\cite{redmark}.


\section{Security Analysis of RDMA Protocols in NVMe-oF}
\label{sec:rdma}

This section analyses the security of RDMA-capable devices and protocols that are used by the NVMe-oF protocol. 
All attacks on RDMA have been discovered during the security analysis of NVMe-oF protocol and its implementations, and have been successfully leveraged to compromise the security of the NVMe-oF protocol. For example, an attack allowing to inject packets into an RDMA connection enables injection into an NVMe-oF connection. Nonetheless, all the attacks can be successfully applied to other RDMA-enabled systems and protocols. Therefore, this section does not solely focus on the NVMe-oF protocol but instead discusses the security of connections by RDMA-enabled applications.
The discovered vulnerabilities and attacks on RDMA have been analyzed for various network controllers supporting RoCE and  InfiniBand protocols.

\begin{table}[t]
\setlength\tabcolsep{0.5pt}
\centering
\resizebox{1.0\linewidth}{!}{%
 \begin{tabular}{@{}p{2.1cm}x{1.0cm}x{1.0cm}x{2.0cm}x{1.8cm} x{2.0cm}@{}}
 \toprule
  \multicolumn{1}{c}{ }& \multicolumn{3}{c}{Injection into} & User access &  Implemen- \\
  \cmidrule{2-4}
  &    RoCE &  IB & IPsec in \ta & in \ta & tation\\ 
 \midrule
 ReDMArk~\cite{redmark} & \cmark & \xmark & \xmark &  \xmark & sockets \\
 \name &  \cmark &  \cmark &  \cmark & \cmark & RDMA verbs \\
 \bottomrule
\end{tabular}}
\caption{Comparison of injection tools. IB - InfiniBand.}
\label{tab:comparison}
\end{table}

\subsection{Packet Injection}
\label{sec:inject}
First, we analyze attacks that allow packet injection into InfiniBand-based protocols including RoCE and native InfiniBand. Compared to existing injection tools~\cite{redmark}, our packet injection attack is feasible without administrative privileges assuming that the adversary is located on the same machine as the victim (see Table~\ref{tab:comparison}.).

In the case of conventional TCP sockets, the operating system prevents packet injection into local connections by not exposing ``raw'' sockets to unprivileged users and isolating sockets belonging to different processes. As RNICs do not offer privileged ``raw'' endpoints, it was considered impossible to forge native InfiniBand packets. We show that any user can inject packets into any connection created on a local RNIC independent of whether it has been created from user- or kernel-space, bypassing security mechanisms of the operating system and its kernel. The implemented injection tool can also successfully inject packets into IPsec over RoCE channels in the \ta model, thereby impersonating even secured RDMA connections, which was not possible with 
the ReDMArk~\cite{redmark} tool.



\subsubsection{Core Vulnerabilities}
We below describe core vulnerabilities found in the InfiniBand architecture and its implementations that allow an attacker to successfully inject packets.

\textbf{Lack of sanity checks during connection creation.}
Users can create an RDMA endpoint without relying on the RDMA connection manager, by directly using the RDMA verbs library (to which we further denote to as \emph{native RDMA connection establishment}).
This approach does not communicate any messages, because it expects that an application receives connection parameters through other means of communication (e.g., using an out-of-band TCP connection). To establish an RDMA connection, each host manually creates an RDMA endpoint with information about the destination host, such as its connection identifier, routing information, and states for incoming and outgoing packet counters. 

Unfortunately, none of the tested hardware providers of RNICs performs basic sanity checks during the creation of an RDMA endpoint using native RDMA connection establishment. This allows creating and using multiple RDMA endpoints that target the same remote RDMA endpoint (i.e., with the same destination QPN) but have different local identifiers. Therefore, two different processes on a host can communicate to the same destination RDMA endpoint. This allows an attacker to create an almost identical copy (except for the source QPN) of a victim's existing connection and use it for communication to the same RDMA endpoint. Given that the creation of an RDMA endpoint using the native RDMA interface does not involve any network communication, the detection of such an impersonation attempt is limited.

\begin{figure}[t]
\centering
\includegraphics[width=1\linewidth]{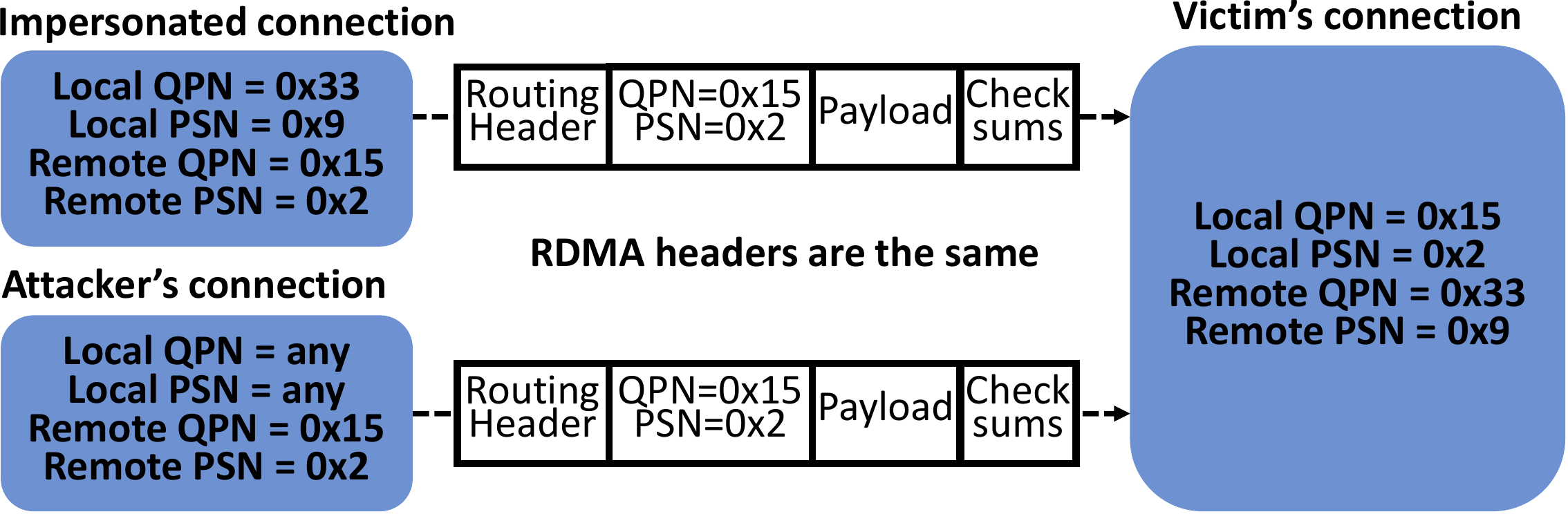}
\caption{An adversary creates an RDMA endpoint with the same destination as the impersonated endpoint, making its packets indistinguishable from the victim. PSN - packet sequence number. QPN - RDMA connection identifier.}
\label{fig:inject}
\end{figure}

\textbf{Source QPN is not contained in RDMA packets.}
InfiniBand-based protocols do not contain the source connection identifier (QPN) in the packet, but only include the destination identifier in the base transport header (see Figure~\ref{fig:overview}). This design choice is based on the fact that RDMA connections are point-to-point channels. Thus, the source QPN is communicated to the receiver when the RDMA connection is established and then stored in a connection table on the receiving endpoint. 

Even though this design choice does not seem like a vulnerability
on its own, it can be combined with the previous
design flaw, allowing an adversary to mimic an existing connection with a different source QPN, to inject packets into the existing connection without
administrative privileges. 
The fact that the source QPN is not included in RDMA packets makes the packets sent by an adversary using its forged connection indistinguishable from regular RDMA packets from the perspective of the remote RDMA endpoint (see  Figure~\ref{fig:inject}). 
Therefore, the combination of these two design flaws enables packet injection into any local RDMA endpoint without any administrative privileges.

\subsubsection{Implementation}\label{sec:packetqpn}
We implemented the packet injection attack using the native InfiniBand interface that uses the RDMA verbs library to manually build a connection endpoint.
For that, the attacker is required to know information about the victim's endpoint such as the network port address, its connection identifier (QPN), and its current packet sequence number (PSN).

The adversary can obtain this information as follows:
\begin{itemize}[nolistsep, leftmargin=*]
    \item The destination network port address is considered public information (e.g., it is the IP address of the targeted RNIC for RoCEv2).
    \item The connection identifier is a 24-bit number generated by the RNIC. However, the generators of connection identifiers in all tested RNICs have shown to be flawed (see Table~\ref{tab:qpn}). The connection identifiers are assigned sequentially and the devices use a static initialization value after a reboot, allowing the attacker to guess a valid connection identifier.
    \item Given that the attacker is co-located with the targeted endpoint and shares the same RNIC, it could create its own RDMA endpoint and gain information on previously established connections based on the identifier assigned to its connection.
    Similarly, if the attacker can legitimately connect to an application on the target machine, it can query the remote connection identifier of the connection with the application.
    Otherwise, the attacker could attempt to start its attack after a scheduled reboot and injecting packets with the identifier that is larger than the static starting seed. We tested 31 different RNICs of five models listed in Table~\ref{tab:qpn}, and the starting seed was in the range from 0x10 to 0x600.
    \item Since the starting packet sequence number is randomly generated by the RDMA connection manager  and an injection of packets with wrong sequence numbers does not affect the victim's connection, we suggest to enumerate all its states (24 bits).
\end{itemize}

\textbf{Injection under the \ta model.}
Using the discussed vulnerabilities, the attacker can impersonate an endpoint sitting on the host and perform the injection into its connections. To verify this, we implemented a tool that does not require administrative permissions and can be run by any user. We have tested the tool on InfiniBand and RoCE networks and successfully managed to inject RDMA requests. Interestingly, we could inject at most 128 packets from an attacker's connection, as it never received acknowledgments from the target endpoint and, therefore, could not make further progress.
Thus, our tool creates an RDMA connection for each set of 128 requests. It is worth mentioning that connection initialization is an expensive procedure that takes around seven microseconds. Thus, the naive implementation of the attack tool that uses only one connection is limited in throughput by the connection initialization. For example, to send $2^{24}$ packets the naive implementation requires about 124 seconds. 

\begin{table}[t]
\setlength\tabcolsep{0.5pt}
\centering
 \begin{tabular}{@{}p{3.85cm} x{2.0cm} x{2.4cm}@{}}
 \toprule
 RNIC model &    Static init. &  Sequential QPN \\ 
 \midrule
Mellanox X3 MT27500 & \cmark  & \cmark  \\ 
Mellanox X5 MT27800 & \cmark & \cmark  \\ 
Mellanox X6-Dx MT28841 & \cmark & \cmark  \\ 
Broadcom BCM57414  & \cmark & \cmark  \\ 
Broadcom BCM58802 & \cmark & \cmark  \\ 
\bottomrule
\end{tabular}
\caption{Analysis of the generators of connection identifiers (QPNs) on tested devices. The same RNIC models can have different starting state, however, all tested devices had a starting state that was smaller than 0x600.}
\label{tab:qpn}
\end{table}

As the packet sequence number is only 24 bits in size, a successful injection of $2^{24}$ packets completely enumerates all its states and resets the counter to the same state as before the injection, thereby making the attack unnoticeable for the impersonated endpoint. 
Thus, fast injection of  $2^{24}$ packets is a desired capability for a successful attack. To have faster injection the attack tool can try to create $2^{17}$ endpoints for each starting packet counter and then start injection from the created endpoints. However, this approach is not always possible as RDMA drivers introduce the limit on created endpoints, which is typically lower than $2^{17}$. Thus, we propose to create all available endpoints and reuse them to inject  $2^{24}$ packets. We first send 128 requests from each created connection once, and then we reset the state of some endpoints to continue the injection. This allows us to enumerate all packet sequence numbers within 1.6 seconds on our Mellanox X6-Dx 25G RNIC, which is an 8x improvement over the ReDMArk injection tool~\cite{redmark}.


\textbf{Injection under the \tb model.}
Injection under the \tb model is performed using the same tool as in the model \ta. However, as was mentioned in Section~\ref{sec:background}, the RNIC matches the destination RDMA connection identifier with the routing header, making packets sent by an unprivileged  user from a remote host dropped because of a mismatch in the routing header. 
Thus, the adversary also needs to change the address of the local network port. For RoCE networks, the attacker can use \textit{ifconfig} tool to assign the IP address of the impersonated endpoint to a local RDMA device. For InfiniBand networks, the attacker needs to change InfiniBand local identifier (LID) of the network port, which is assigned by a subnet manager. The subnet manager is a management application of InfiniBand network that assigns addresses to all RDMA devices, and it cannot be configured with the ifconfig tool.
Even though it is generally believed that the LID can be assigned only by the subnet manager, we found out that the \textit{ibportstate} tool from the \textit{infiniband-diags} package allows privileged users to change the configuration of the RNIC (see Figure~\ref{code:lid}). 
This change is not permanent as the subnet manager detects the changes and reassigns the correct LID. 
However, the update intervals can take several seconds, which is enough to create a device context and start the injection. On our InfiniBand cluster, we were always able to change the LID of any port with the \textit{ibportstate} tool and inject packets with the required routing header. 

\begin{figure}[t]
 \begin{minted}
[
frame=lines,
framesep=0.3mm,
baselinestretch=0.1,
fontsize=\footnotesize,
]
{java}
$ sudo ibportstate -D 0 1 lid 0x4 active -d && ibv_devinfo
\end{minted}
\caption{An example of the command to set 0x4 as InfiniBand local identifier (LID) of the first RNIC, and to print it. }
\label{code:lid}
\end{figure}

\textbf{Injection into IPsec over RoCE}\label{sec:ipsec}
The IPsec over RoCE protocol~\cite{ipsecroce-slides} is designed to bring secure channels to RoCE packets. IPsec is an internet layer protocol that encapsulates RoCE packets and protects them. It means that IPsec secures RDMA-enabled applications at the IP layer, which allows deploying  all existing applications with no code changes as the security policies are configured by an administrator for each IP-IP path. The security keys are known only to the administrator, thereby preventing users to sniff and read packets of other users.

As the security policies are installed based on source and destination IPs, RDMA connections created on the same path share the same secure context (i.e., IPsec policy). In other words, if we have two users sharing the same RNIC, they both share the same IPsec policies and the operating system must provide isolation between connections created by two different users. 

We have tested our spoofing tool under the model \ta with enabled IPsec over RoCE. We configured a secure IPsec channel in transport mode between two Mellanox X6-Dx cards~\cite{ipsecroce-setup}. As IPsec rules are the same for two RDMA connections of different users on the same IPsec-enabled RNIC, the IPsec packets are also indistinguishable by the receiving RNIC. The problem is that RoCEv2 packets encapsulated into IPsec packets do not use UDP ports to recognize the sender and the receiver (the source UDP port is any random number and the destination UDP port is a fixed number reserved for RoCEv2 protocol). RoCEv2 packets fully rely on the base transport header of the InfiniBand architecture, which is opaque to IPsec, to recognize the receiver and the sender using only the destination RDMA connection identifier.
This fact allows an adversary to inject packets into RDMA connections that are protected by IPsec using our tool.

We have also tested the injection tool from ReDMArk~\cite{redmark} and it was not able to successfully inject a packet into an RDMA connection protected with enabled IPsec, as their tool is based on raw sockets requiring administrative permissions. IPsec policies could  detect and drop these packets.

\subsubsection{Mitigation}
To mitigate the injection under the \ta model, the RDMA providers and the specification should address the two vulnerabilities that we discovered. 
A simple solution is to make a firmware update that disallows users to have two local connections with the same destination. However, it may increase the connection time, as the RNIC would need to scan destinations of all connections. What is more, such a naive solution may even reveal to the adversary all remote destinations, as the adversary could locally probe possible remote identifiers. 
A more robust approach is to add a four-byte header (RDMA packets must be 4-byte aligned) that would include the source QPN (24 bits) to the packet format, allowing the RNICs to distinguish packets from different connections. Alternatively,  a secure transport should be introduced to the InfiniBand architecture, as proposed by sRDMA~\cite{srdma}. However, both solutions would require changes to the architecture. Alternatively, one can leverage programmable data planes to modify RDMA packets to enable source authentication, as proposed by Xing et al.~\cite{bedrock}. Their authentication tool can mitigate our injection attack, but it requires specialized programmable network controllers and switches, that are not always available.  

\subsection{Slow Down using Congestion Control}\label{sec:slowdown}
In this section, we show how a privileged attacker can deceive RDMA endpoints into thinking that they experience congestion, forcing them to slow down.

\subsubsection{Vulnerability}
InfiniBand-based protocols support congestion control to prevent packet drops because of bursted traffic. 
Switches of an RDMA network mark packets contributing to the congestion by setting a congestion bit in the transport header. 
%
The congestion notification is carried through to the target, which generates a \emph{congestion notification packet} that advises the initiator to reduce the injection rate to resolve congestion. 
When receiving a congestion notification packet, the RNIC reduces the rate of injection for the RDMA connection indicated in the packet.

The main vulnerabilitiy is that congestion notification packets are not protected, allowing an attacker to forge them. Forging a congestion notification packet only requires knowing the connection identifier (QPN) of the victim.  The packet sequence number field must be zero, as it is designed to be an out-of-order packet that could be potentially lost. \emph{As a result, anyone in the network can forge a congestion notification packet to disrupt the normal operations of any RDMA-enabled application.}

\subsubsection{Implementation}
\begin{figure}[t]
\centering
\includegraphics[width=0.94\linewidth]{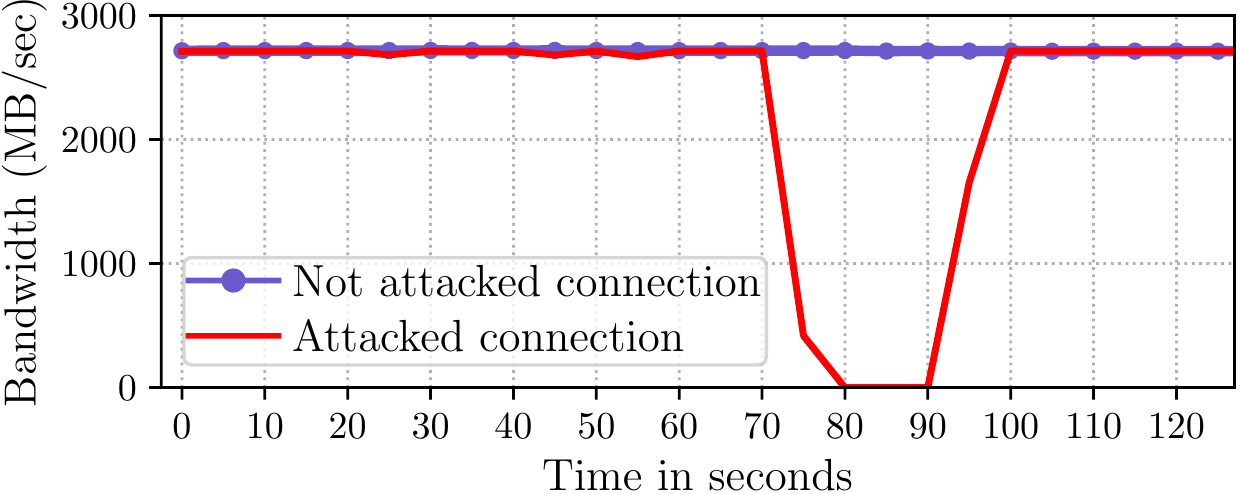}
\caption{Bandwidth of the victim connection during the injection of congestion notification packets. The injection of 10M packets started at second 70 and took 20 seconds. }
\label{fig:inject-write}
\end{figure}

As the RDMA verbs library does not expose congestion control to user space, we cannot use our \name{} tool for injection. Instead, we extended the ReDMArk injection code to inject congestion notification packets for the RoCEv2 protocol. 
To measure the effectiveness of the attack we performed an attack on the bandwidth benchmark of the perftest suite~\cite{perftest}.
We created a victim connection that was used for the uni-directional RDMA write benchmark. 
Then we run our attack tool that injects congestion notification packets targeting the endpoint that issues RDMA writes. Figure~\ref{fig:inject-write} shows the speed of the victim connection before, during, and after the attack. To confirm that the slowdown happens due to the attack on the congestion control, we measure the  bandwidth of the victim connection when we injected congestion notification packets into another connection on that device.
The measurement shows that we significantly decreased the performance of the victim connection from 2713 MB/sec to 0.04 MB/sec. Injection of packets to another connection did not slow down the victim connection.

\subsubsection{Mitigation}
The simplest mitigation technique is to use IPsec over RoCE that drops such spoofed packets. It also would be beneficial to have a secure transport in the InfiniBand architecture. Nonetheless, a proposal for a secure RDMA transport, sRDMA~\cite{srdma}, does not address the congestion control and does not state whether it is supported. We assume that sRDMA does not support pure congestion notification packets as they do not have a packet counter required for salting message authentication codes, preventing sRDMA to uniquely authenticate each congestion notification packet. On the other hand, the InfiniBand architecture allows piggybacking congestion notification with RDMA responses that are secured by sRDMA. 
Thus, for a future secure transport with congestion control, InfiniBand-based protocols would need to transfer congestion notifications within ordered packets such as acknowledgments or make specialized ordered secure congestion notification packets. 



\subsection{Attacking RDMA connection manager}\label{sec:rdmacm} 
In this section, we show how an unprivileged attacker can deceive RDMA endpoints into thinking that their connections are disconnected. For that the attacker exploits vulnerabilities in RDMA connection manager allowing the attacker to spoof disconnect requests, forcing the user to receive a falsified disconnect event. 

Applications use RDMA connection manager to send connect and disconnect requests, that contain secret connection keys for proving authenticity of requests. Even though these secret connection keys are managed by the corresponding kernel module and are not known to applications, we found a vulnerability in their generator, allowing an attacker to acquire secret keys for building falsified disconnect requests.

\subsubsection{Vulnerability}
\textbf{Lack of request filtering.}
The RDMA connection manager kernel module works via unreliable RDMA connections (they are similar to UDP sockets) with a special reserved connection identifier  (QPN is equal to 1).
As it is unreliable, it is able to receive packets from any endpoint and process them after verifying the included secret connection keys. 
Even though unreliable connections, unlike reliable connections, include information about the source (i.e., QPN of the sender), the kernel module does not check it and processes all received messages regardless of the sender. \emph{So any RDMA application, even unprivileged, can send a message to the kernel module of the connection manager.}

\textbf{Weak key generator.}
To build a disconnect request the adversary needs to provide only three values: 
the connection identifier (QPN) that it wants to disconnect, the secret connection key of the initiator, and the secret connection key of the target. The connection key is a 32-bit value assigned by the connection manager kernel module for each local connection, and user applications are not privileged to know it. 

%

\emph{We argue that the generator of connection keys is weak and has low entropy.} 
The full algorithm for the generator is listed in Figure~\ref{fig:rdmacm}. As we can see the kernel module gets a random 32-bit starting seed when it is loaded, and then XORs it with sequential identifiers.
Due to the nature of the algorithm, the difference between the two keys is usually in the least significant byte and in most cases is one bit. Therefore, if the adversary can guess a recent key it can guess keys of other recent connections or at least enumerate them without knowing the starting seed. 

\begin{figure}[t]
 \begin{minted}
[
frame=lines,
framesep=0.3mm,
baselinestretch=0.1,
fontsize=\footnotesize,
]
{cpp}
//on loading the RDMA connection manager kernel module
uint32_t seed = get_random_bytes(sizeof(uint32_t));
uint32_t local_key_state = 0;

//to assign hidden connection keys
uint32_t get_key(){
    return seed ^ (local_key_state++);
}
\end{minted}
\vspace{-0.25cm}
\caption{A pseudo-code (in C language) showing how the RDMA connection manager assigns keys to connections. }
\label{fig:rdmacm}
\end{figure}

\subsubsection{Implementation}
We extended our injection tool to use unreliable connections to generate disconnect requests. In our first experiment, we checked whether a client with user permissions can send a disconnect request from user space and disconnect a remote reliable connection. For that, we captured  (using tcpdump~\cite{tcpdump}) the connections keys and the target connection identifier (QPN) from the network to generate the correct disconnect request. We could do it as the RDMA connection manager communicates all connection parameters in plain text.
\emph{Our experiments showed that any unprivileged endpoint within the network can falsify a disconnect request and disconnect any reliable connection, even sitting on a machine different from the connection endpoints (i.e., location of the \tb model, but capabilities of the \ta model).}
Thus, this attack can be performed even with IPsec over RoCE enabled and from any location on the network without administrative privileges.
Interestingly, for the tested devices the reception of the disconnect request does not break or destroy an RDMA connection and only works as a notification to the user. However, all applications, including NVMe-oF implementations, are implemented to trust these notifications and destroy connections.

\textbf{Tool for finding local secret connection keys.}
To perform this attack without sniffing capabilities we propose a technique for finding secret connection keys of local RDMA connections. 
\emph{Our tool allows the attacker to find secret keys of victim connections on machines into which the adversary can login. }

The tool creates two RDMA connections in one process and connect them using the RDMA connection manager. As a result, the connections will get two secret RDMA connection keys that are likely to have a difference in 1 bit because of the used generator (see Figure~\ref{fig:rdmacm}). 
Using this information, the adversary can locally probe disconnection requests to detect the state of the connection key generator.
As the connection identifier (QPN) is known, the adversary only needs to enumerate 32 bits (the first key) plus 3 bits (the flipped bit in the first byte of the second key) to find both hidden keys. This process can fail if the difference was in more than 1 bit, but the attacker can retry it again.  Our tool manages to find the keys within one hour as it  generates 10M requests per second on our Mellanox 6X-Dx 25G RNIC. 

After that, the adversary can guess the initial random seed of the local RDMA connection manager. If the machine was recently booted, the attacker can have a close guess on the number of generated keys. 
For example, by exploiting the fact that RDMA connection identifiers (QPNs) are generated sequentially from a static starting state. 
Thus, the attacker can guess the number of created RDMA connection manager contexts by calculating the difference between the starting and the current states of the QPN generator. By having the guess, the adversary can simply find the seed by XORing the resulting value with a recently found identifier.

\subsubsection{Mitigation}
To improve the security of the RDMA connection manager, we propose to process requests only from the reserved connection for the connection manager. It will prevent the injection from unprivileged users, as the reserved connection can be created only from the kernel. 
Interestingly, we performed a similar experiment with the IP-over-IB protocol~\cite{ipoverib}, which uses another reserved connection (the QPN is equal to 0x208) for encapsulating IP packets into unreliable RDMA packets,  but it checks the source QPN, thereby removing a huge attack surface.  

Nonetheless, for RoCE networks, an adversary with administrative permissions can forge a packet with any source connection identifier (QPN), making the attack still possible under the \tb model. 
Thus, we propose applications to verify disconnect requests.
 Developers can make use of payloads that can be sent with a disconnect request to distinguish trusted disconnect requests from untrusted at the application level. 
 As a connection can be broken due to application failure, we propose to verify all untrusted disconnect requests by sending a challenge message to a remote endpoint using challenge handshake authentication protocol~\cite{chap}. If the remote endpoint does not reply after a series of retries, the connection can be closed.

\subsection{RDMA Connection Exhaustion Attack}\label{sec:exhaust}
In this section, we show how a user with no administrative privileges can prevent other applications from opening RDMA connections. The attack can be performed by any user and affects all RDMA-enabled applications. 

\subsubsection{Vulnerability}
RDMA drivers introduce limits on the number of open RDMA connections to operating systems.
Unlike Linux, which limits the number of open TCP connections by each user, RDMA limits are system-wide. As a result, all RDMA-enabled applications, including privileged applications running in the kernel, share the same limit on the number of open connections. This vulnerability allows any local user to exhaust this limit and prevent other applications to have new connections. This applies also to other RDMA resources such as completion queues and memory regions, but here we focus on connections only. 

\subsubsection{Implementation}
We implemented a tool that locally creates RDMA connections with the RDMA verbs library until it reaches the limit. After running the tool  we attempted to create an RDMA connection from user and kernel spaces. Our experiment showed that our tool successfully blocks new connections for user space applications as well as kernel space applications. 

\subsubsection{Mitigation}
To allow administrators to manage RDMA resources, the integration of RDMA and its limits into current operating systems should be improved.  
After contacting some RDMA providers, they described an approach allowing administrators to introduce user limits to RDMA resources called \emph{``RDMA controllers''}~\cite{rdma-controller}, which is a component of Linux control groups. RDMA controllers are not widely used and not many administrators know about their existence. 
We want to mention this technique here to inform other administrators, as it is not mentioned in the RDMA aware programming manual~\cite{manual} and we find this useful tool insufficiently promoted.

\subsubsection{Comparison to other Exhaustion Attacks}
ReDMArk~\cite{redmark} proposes remotely exhausting RDMA connections available to applications, prohibiting them from opening new connections.
Even though they target the same vulnerability, they are different and require different mitigation techniques. 
ReDMArk's attack is performed from a remote location by a user which is able to establish many RDMA connections with an RDMA-enabled system to prevent the system from opening connections with other clients. Our attack does not require an ability to connect to the RDMA-enabled system, which makes it applicable to NVMe-oF applications under the \ta model. Regarding mitigation, our attack can be mitigated by the RDMA controllers, whereas RDMA controllers make ReDMArk's attack even more viable. ReDMArk's attack is mitigated using authentication of clients or by limiting the number of open connections from the application, thereby making the attack impractical for NVMe-oF targets, which already prevent unauthorized connections and  monitor open connections.

\section{Security Analysis of the NVMe-oF Protocol}\label{sec:nvmeof}
In this section, we analyze the security of the NVMe-oF protocol and its implementations.
First, we give a background on NVMe-oF requests and the security mechanisms of NVMe-oF.
Then, we show how the attacks from Section~\ref{sec:rdma} can be effectively applied to NVMe-oF applications. Finally, we propose mitigation techniques that will improve the security of disaggregated storage protocols.

\subsection{NVMe-oF and its Implementations}
The NVMe-oF protocol allows the execution of NVMe commands on a remote solid-state storage device or system
over RDMA-capable networks.
For this purpose, it transfers the NVMe commands and responses between the current and the target host by encapsulating them into the payload of RDMA messages. 
THe NVMe protocol differentiates between control-plane and data-plane capsules. The former are transferred
using RDMA send, whereas the latter are transferred using one-sided memory accesses such as RDMA read and write operations.
 

\subsubsection{Request Types}
\begin{figure}[t]
\centering
\includegraphics[width=1\linewidth]{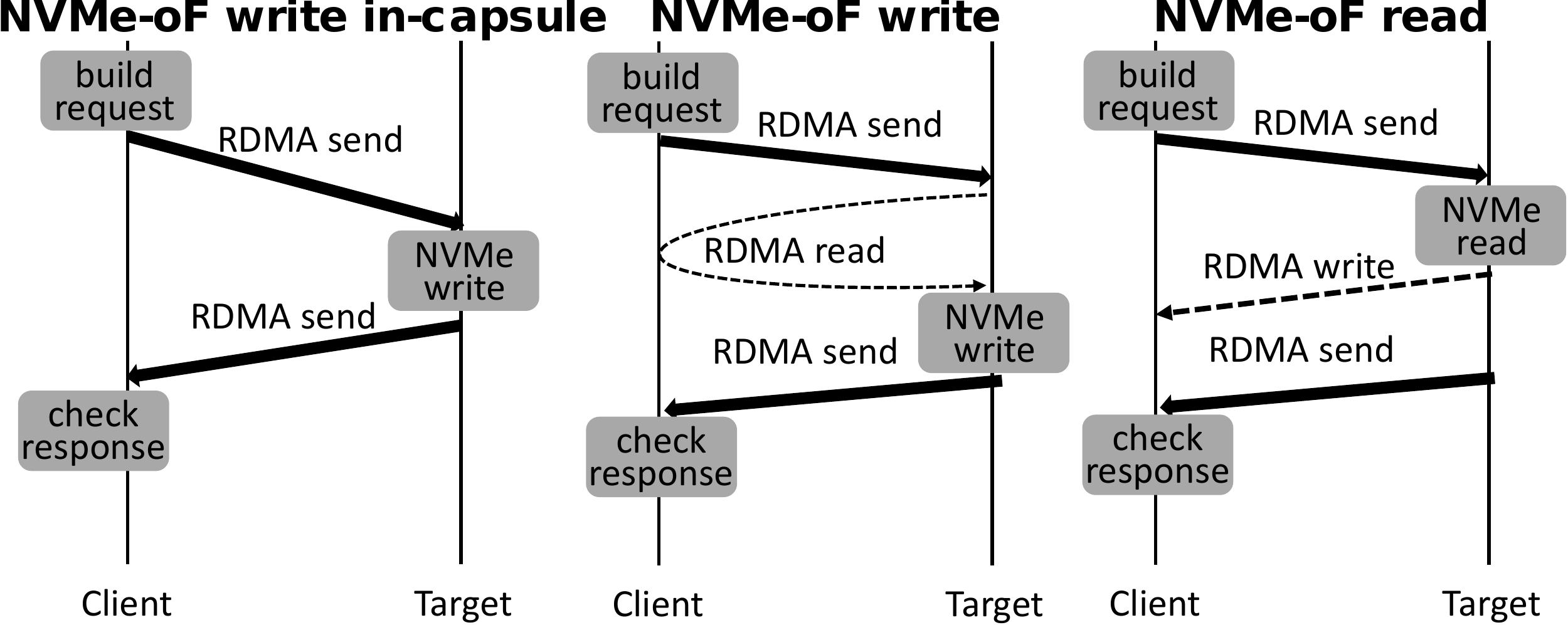}
\caption{NVMe-oF requests and their implementation over RDMA connections.}
\label{fig:nvmeof}
\end{figure}

\textbf{NVMe-oF Write in-capsule.}
Write requests with a small payload size (usually less than 4096 bytes) can be sent as a single message via a single RDMA send request. Besides the command data, the request capsule contains basic information such as a destination block address, the command identifier, command parameters, etc.
Upon receiving the capsule, the NVMe-oF target checks the correctness of the request and writes data to the local NVMe device as illustrated in Figure~\ref{fig:nvmeof}. Finally, the target sends a response to the client, indicating the completion of the request.
 
\textbf{NVMe-oF Write.}
If a client wants to send a large payload, the request will contain the information allowing the NVMe-oF target to fetch that data from the client using an RDMA read operation. After receiving the request, the target reads the data from the client. After the data is read and written to local NVMe, the target sends a response indicating the completion of the NVMe-oF request. 

\textbf{NVMe-oF Read.}
To read data from a remote NVMe device, the client sends an NVMe-oF read request that contains the information where the target should write the data. The target will read the data from the disk and then write it to the buffer of the client using an RDMA write request. As the RDMA write is a ``silent'' one-sided operation, the target sends a response via an RDMA send to indicate the completion of the request. 

\subsubsection{In-band Authentication}
\label{sec:inband}
The NVMe technical working group released a specification~\cite{nvme,nvmeof} defining an in-band authentication protocol for NVMe protocols called the \emph{DH-HMAC-CHAP}, which provides bidirectional authentication between a client and a target using the challenge handshake authentication protocol (CHAP). The DH-HMAC-CHAP protocol is an enhanced CHAP protocol with hash-based message authentication code and augmented with an optional Diffie-Hellman (DH) exchange.
The DH-HMAC-CHAP is used to authenticate connections (i.e., between clients and targets), but cannot be used to authenticate individual NVMe-oF requests. The current specification suggests to authenticate the endpoints using DH-HMAC-CHAP and then use TLS or IPsec to provide authentication and encryption for the communication channels.

The core problem of the proposed security extension is that it does take into account that RDMA connections do not offer a secure transport like TLS~\cite{rfc8446}. 
The in-band authentication can only be used by TCP connections and targets a TCP extension to the NVMe-oF protocol, called NVMe-oF/TCP.
Thus, the NVMe-oF protocol fully relies on IPsec over RoCE to provide secure communication to RDMA connections.

\begin{table*}[t]
\setlength\tabcolsep{0.5pt}
\centering
\begin{threeparttable}
\resizebox{0.84\linewidth}{!}{%
 \begin{tabular}{@{}p{4.9cm}x{1.2cm}x{1.2cm}x{1.2cm}x{1.2cm}x{1.2cm}x{1.2cm}x{4.2cm}@{}}
 \toprule
 \multicolumn{1}{c}{ }& \multicolumn{3}{c}{Threat Model \ta} & \multicolumn{3}{c}{Threat Model \tb}  \\
 \cmidrule{2-4} \cmidrule(l){5-7}
 Attack &   None  & In-band & IPsec &   None  & In-band & IPsec & Effect \\ 
 \midrule
 Spoof NVMe-oF request & Yes & Yes & Yes  & Yes & Yes & No   & Execution of falsified request \\
 Spoof NVMe-oF response & Yes & Yes & Yes &  Yes & Yes & No  & Early termination \\
 Corrupt memory &  Yes\tnote{1} & Yes\tnote{1} & Yes\tnote{1} & Yes\tnote{1} &  Yes\tnote{1} & No & Use of falsified data \\ \midrule
 Exhaust RDMA connections &  Yes\tnote{2} & Yes\tnote{2} & Yes\tnote{2} & No & No & No  & Connection failure \\
 Spoof congestion notification packets &  No\tnote{3} & No\tnote{3} & No\tnote{3} & Yes & Yes & No  & Connection slowdown\\
 Spoof RDMA disconnect  &  Yes & Yes & Yes & Yes & Yes & Yes & Disconnection  \\
 Spoof invalid packet~\cite{redmark} &  Yes & Yes & Yes & Yes& Yes & No & Disconnection  \\  
  \cmidrule{2-7}
   \multicolumn{1}{c}{ }& \multicolumn{6}{c}{Feasibility of the attacks on NVMe-oF}\\
  \bottomrule
\end{tabular}}
\begin{tablenotes}
\footnotesize
\item[1]  Linux kernel uses fast memory registrations with invalidation, which increases the complexity of the attack.
\item[2]  Can be mitigated with RDMA Controller~\cite{rdma-controller}.
\item[3]  Injection of congestion notification packets is possible only for RoCE with administrative permissions.
\end{tablenotes}
\end{threeparttable}
\caption{Analysis of attacks on the NVMe-oF protocol depending on enabled security mechanisms (None, In-band, or IPsec).}
\label{tab:attacks}
\end{table*}

\subsubsection{IPsec over RoCE}
Since the previously discussed in-band authentication protocol neither offers message authentication nor encryption for NVMe-oF, the NVMe technical working group currently proposes to employ IPsec over RoCE~\cite{ipsecroce-slides}. IPsec over RoCE provides data secrecy for the NVMe-oF protocol deployed over the RoCEv2 protocol. However, IPsec is not implemented for InfiniBand networks as it is not an IP-based protocol, making InfiniBand interconnects vulnerable to attacks under the threat model \tb.
In addition, as discussed in Section~\ref{sec:ipsec}, IPsec does not provide isolation between RDMA connections of different users: applications that use the same IPsec-enabled network interface share its security policies, making them vulnerable to attacks under the threat model \ta (see Section~\ref{sec:inject}). Therefore, \emph{IPsec is not sufficient to offer a secure transport for the NVMe-oF protocol.}

\subsubsection{NVMe-oF Implementations}
\textbf{SPDK.} 
The Storage Performance Development Kit (SPDK)~\cite{spdk} provides a set of tools and libraries for writing high-performance, scalable, user-mode storage applications. SPDK is a  user space library for accessing NVMe-enabled devices and is already used by many database and storage systems~\cite{crail, crailkv, polarfs, nvme-dbms, reflex, hive}.
In other words, device driver code runs at the user level, avoiding kernel context switches and interrupts. SPDK library offers an NVMe-oF client library and an NVMe-oF target application for RDMA connections. In addition, SPDK exploits hugepages~\cite{hugepages} to reduce memory translation overheads that are present in RNICs~\cite{farm}. 

\textbf{Linux kernel.} 
The Linux NVMe driver~\cite{nvme-linux} is included as part of the Linux kernel. NVMe-oF clients and targets are implemented as kernel modules that run with administrative permissions. Unprivileged Linux users cannot directly interact with NVMe devices as NVMe provides block-level access to the disks. Therefore, users interact with NVMe devices through high-level interfaces (e.g., a filesystem) that translate block requests to NVMe-oF requests. In summary, the Linux NVMe driver is implemented exclusively for kernel space modules and applications.

\textbf{SPDK and Linux NVMe driver comparison.}
The main difference between SPDK and the Linux NVMe driver is in the management of RDMA-accessible memory. In general, all memory that can be accessed by an RNIC must be registered, and any remote user that wants to access memory should include a corresponding access token (memory key) in its one-sided RDMA request. 
Memory registration is different for kernel and user space. 
The kernel can efficiently dynamically register memory with fast memory registration~\cite[p.~381]{fastmr}, which allows the kernel to efficiently register any memory region provided by upper layers. 
Since the kernel is privileged to exploit the fast memory registration, the NVMe-oF driver also dynamically deregisters the memory after it was used by a remote client (using ``memory invalidation'' capabilities of RNICs). 
Such an approach improves the security of NVMe-oF kernel modules as the memory is only RDMA accessible during a short time interval. On the other hand, SPDK cannot efficiently register/deregister memory and therefore registers the memory once and reuses a single registration (i.e., a single memory key) for all RDMA-accessible memory.

\subsection{Attacks on NVMe-oF}\label{sec:nvmemem}
In this section, we describe how the attacks described in Section~\ref{sec:rdma} can be used to circumvent security mechanisms of NVMe-oF. Overall, we propose seven different attacks on the NVMe-oF protocol that are summarized in Table~\ref{tab:attacks}.

\subsubsection{Spoofing of NVMe-oF Requests}
RDMA packet injection allows an attacker to inject an RDMA send request that contains an NVMe capsule. This is possible because NVMe messages are not authenticated and the attacker can guess the connection identifier of the NVMe-oF target as it is usually launched during the boot process. Furthermore, the packet sequence number can be enumerated within 2 seconds using our injection tool.

We have tested this attack on NVMe-oF Write in-capsule requests and could trick both the Linux driver implementation and SPDK implementation to write forged blocks to the NVM device. 
Consequently, \emph{our NVMe-oF capsule injection allows re-writing any NVMe block stored on a remote disk without administrative privileges,  bypassing security mechanisms of operating and file systems.}

\begin{figure}[t]
\centering
\includegraphics[width=0.94\linewidth]{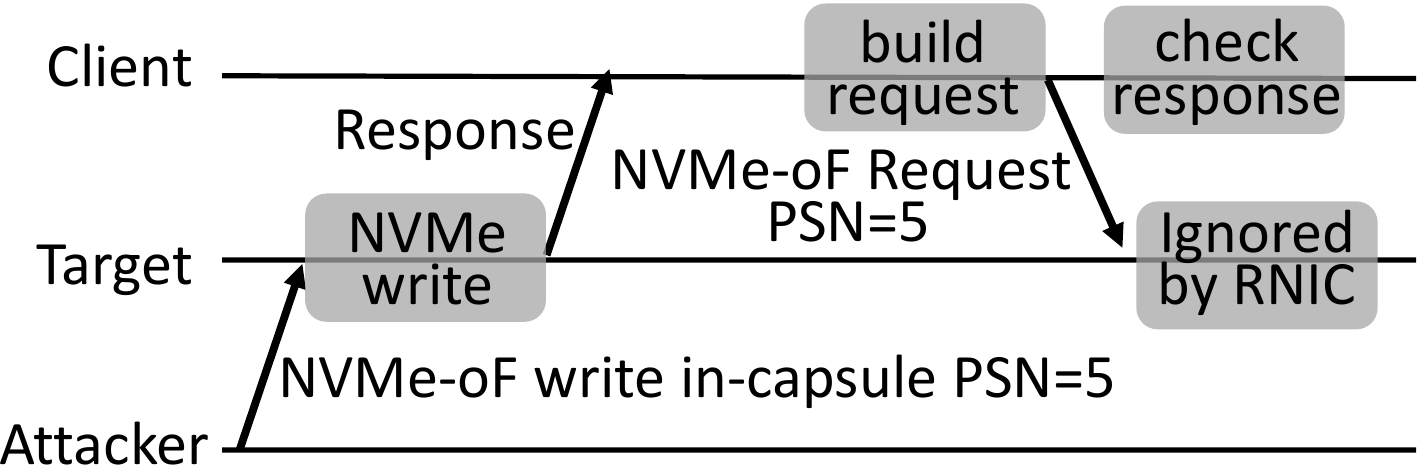}
\caption{Attacking NVMe-oF write in-capsule requests.}
\label{fig:writecapsule}
\end{figure}

Interestingly, the SPDK and Linux kernel implementations behave differently in case of a packet injection attack.
Injection of a single packet does not break the connection for an SPDK client, but always causes a disconnect for a Linux kernel client. This happens due to a de-duplication flow integrated into the RNICs and the way SPDK clients handle responses.
As was discovered in ReDMArk~\cite{redmark}, injection of one packet does not break an established connection as the packet arriving from an impersonated connection is considered to be a duplicate and thus is dropped (due to duplicated packet sequence number (PSN)). This phenomenon is illustrated in Figure~\ref{fig:writecapsule}, where an attacker injects a request that gets executed and responded by the target. When the client sends its request after that, it is dropped by the RNIC due to duplicated PSN. The SPDK client keeps the response for the forged request and treats it as a response for its next request.
The Linux driver implementation, on the other hand, simply ignores the premature response and as a result, does not get any response for its next request. 

\subsubsection{Spoofing of NVMe-oF Responses}
An adversary can spoof the response issued by the target to the client. The reception of a response for a client means that the affected communication buffer has been used and it can be now deregistered or reused for a new request. 
Thus, the injection of NVMe-oF responses to NVMe-oF clients can cause premature memory invalidation for the Linux kernel implementation and premature memory mutation for SPDK.

\begin{figure}[t]
\centering
\includegraphics[width=0.94\linewidth]{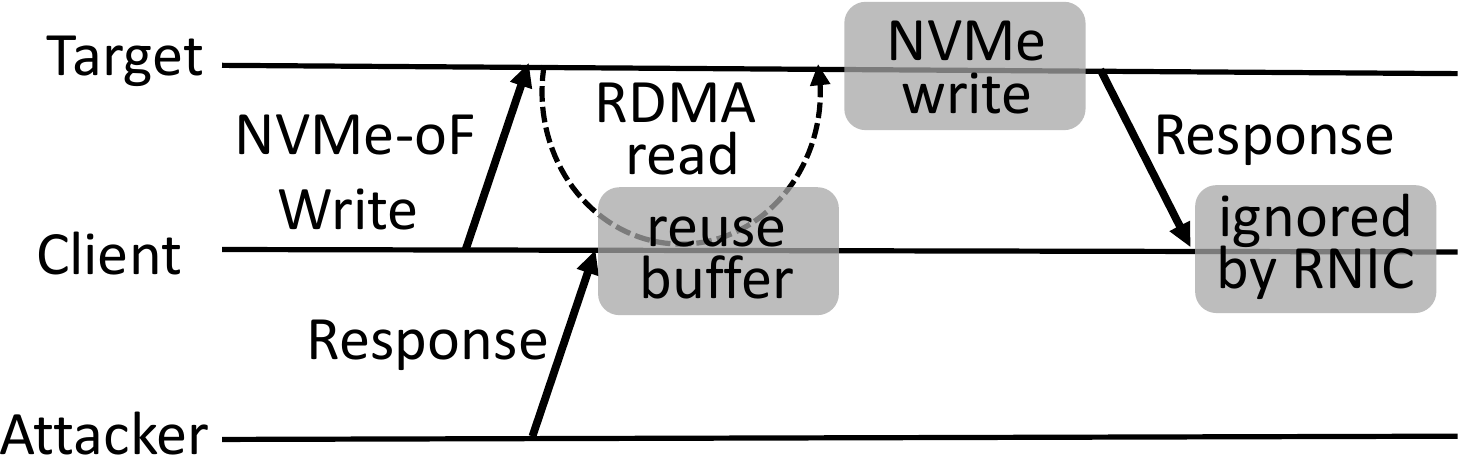}
\caption{Attacking NVMe-oF write requests.}
\label{fig:write}
\end{figure}

We tested this attack for affecting NVMe-of write requests. In the case of  the \ta model, the attack tool was executed on the same machine as the NVMe-oF target. The attack is illustrated in Figure~\ref{fig:write}, where the adversary sends a response right after the RDMA read request arrived from the target. As the client received the response earlier, it may mutate the data before it is fully read by the target, resulting in storing corrupted data into the disk.
Consequently, \emph{our NVMe-oF response injection allows corrupting data on a remote disk without administrative privileges.}

\subsubsection{Memory Corruption using RDMA Write}
An attacker can inject RDMA write requests to change the RDMA-accessible memory of the NVMe client as well as the NVMe target. In theory, an NVMe-oF target should be resilient against this attack as it may register memory only for local accesses, since clients never issue RDMA reads and writes (see Figure~\ref{fig:nvmeof}). Nonetheless, the SPDK target registers memory for remote accesses, making it vulnerable to this attack. The Linux NVMe-oF target module does not have this issue.

 \begin{figure}[t]
\centering
\includegraphics[width=0.94\linewidth]{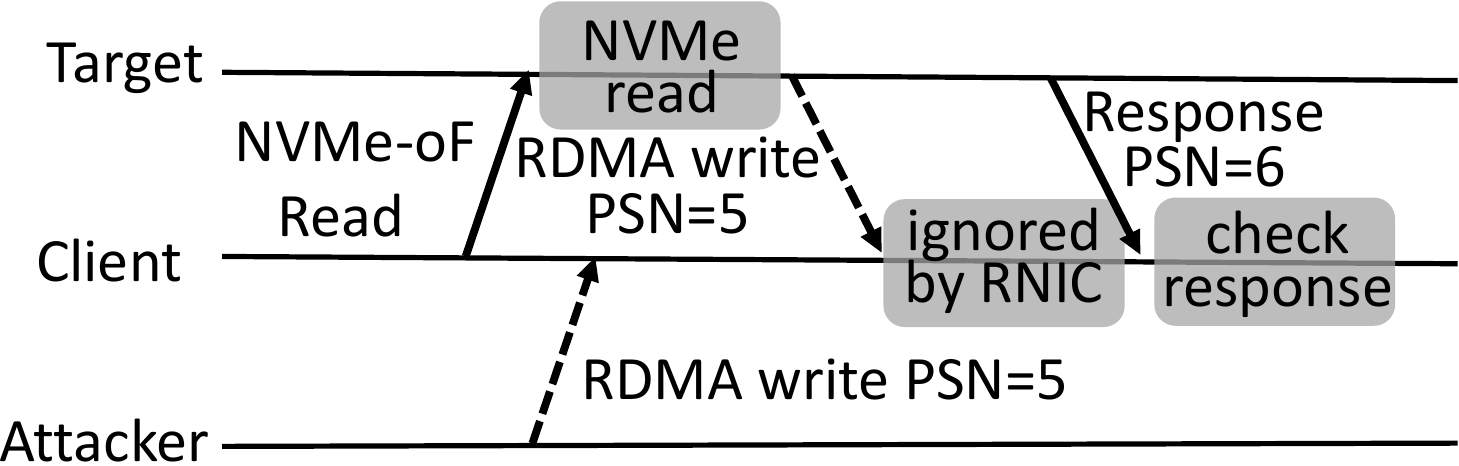}
\caption{Attacking NVMe-oF read requests.}
\label{fig:read}
\end{figure}

Injecting an RDMA write requires the attacker to know a valid memory key (rkey) and the corresponding memory address of the victim endpoint. SPDK clients and targets pre-allocate memory and thus use a single memory key for their communication buffers. Additionally, ReDMArk has shown that memory key generators of existing RNICs are weak and often have problems with static initialization~\cite{redmark}. Thus, some RNICs assign the same memory key to the first memory registration after a reboot. This results in an SPDK application having static predictable keys as they are often loaded at the boot.
The address of memory pools is also known for SPDK as it uses hugepages and by default maps them to a predefined virtual address (0x200000000000). Therefore, an attacker can exploit these vulnerabilities to modify the memory of SPDK applications using RDMA write requests.
 
We tested this attack to alter the data of NVMe-oF read (see Figure~\ref{fig:read}). The attacker issues an RDMA write to the receive buffer of the client application. The client issues an NVMe-oF read to the target, which fails to write the read bytes to the memory of the client as its RDMA write is dropped because of duplicate packet sequence number (PSN). Then the target sends the response, indicating the completion of the operation. As a result, the client observes data completely falsified by the attacker.  Consequently, \emph{this attack allows an unprivileged user to manipulate receive buffers of NVMe-oF clients without affecting the data stored in the remote disk.}

The Linux kernel implementation uses memory registrations that are valid only during a short time interval and uses dynamic buffer addresses that are defined by applications that implicitly use the NVMe-oF kernel module. Therefore, it is mostly resilient to memory corruption attacks and can only be exploited in very specific settings.
To conduct this attack on the Linux kernel implementation, we assume an I/O queue of depth 1 and that the adversary was able to eavesdrop on the last used packet sequence number and the memory key of the last memory request. As the Linux kernel implementation uses fast memory registration (similar to memory windows type 2 in InfiniBand architecture~\cite{infiniband2000infiniband}), the next memory key is equal to the last key plus one. In a zero-knowledge environment with many applications that use many deep I/O queues, the attack becomes more complex but is still considered possible.

\subsubsection{NVMe-oF Connection Slow-Down}
Since NVMe-oF is based on RDMA it is subject to the slowdown attack introduced in Section~\ref{sec:slowdown}. \emph{Our tool allows the attacker to inject congestion notification packets towards the target or the client to slow them down.}
The attacker can guess RDMA connection identifiers of the NVMe-oF applications using the vulnerabilities in number generators discussed in  Section~\ref{sec:packetqpn}. We tested this attack on both NVMe-oF implementations and observed a slowdown as in Figure~\ref{fig:inject-write}.

\subsubsection{Spoofing of RDMA Disconnect request}
NVMe-oF uses the RDMA connection manager to establish connections. Therefore, the attacker can perform the attack described in Section~\ref{sec:rdmacm} to disconnect clients. SPDK and Linux kernel clients automatically reconnect to the target, but the disconnection can cause long periods of unavailability.
Similar to the previous attacks, the attacker can use the fact that often NVMe-oF targets and clients are loaded after the boot process, allowing the attacker to guess the hidden connection keys to spoof a valid  RDMA disconnect request.

\subsubsection{Connection Establishment Prevention}
Using the RDMA connection exhaustion attack discussed in Section~\ref{sec:exhaust}, an attacker can prevent clients and targets from connecting and can be conducted either at the client or at the target machine.
The attack can be performed in combination with forced disconnection attacks. The attacker can disconnect existing connections using the disconnect attacks and follow-up with the exhaustion attack to prevent the target from reconnecting to the client.  

\subsubsection{Forced Disconnection with Invalid Packets}
\label{sec:invalid}
Finally, we apply the attacks proposed in ReDMArk~\cite{redmark} to the NVMe-oF protocol. Most of the proposed attacks can not be applied to NVMe-oF (we discuss the feasibility of the attacks in Section~\ref{sec:related}).
However, the ReDMArk's attack that causes disconnections of connections by injecting invalid RDMA packets is feasible in NVMe-oF. The attacker can inject an invalid RDMA packet that will cause an error on the NVMe-of applications. For example, the attacker can inject an RDMA write with a payload that contains an invalid memory address (e.g., 0x0) and force an existing connection to disconnect. Note that with our injection tool, the injection of invalid packets can be done without administrative permissions, whereas ReDMArk's tool requires them. 

Both SPDK and the Linux kernel are vulnerable to this attack. Similar to previous injection attacks, the attacker needs to know the connection identifier of the victim endpoint and its current packet sequence number to break the connection. Both of these values can either be guessed or enumerated with low effort.

\subsection{Mitigations for NVMe-oF Vulnerabilities}\label{sec:mitig}
The NVMe working group suggests to utilize IPsec and endpoint authentication using the DH-HMAC-CHAP protocol to mitigate the current security shortcomings of the InfiniBand architecture. However, even with the deployment of these mechanisms, not all of the suggested attacks are mitigated (see  Table~\ref{tab:attacks} for a summary).
The problem is that the proposed techniques do not provide an RDMA secure transport and consequently are subject to packet injection attacks in case of a local unprivileged attacker (\ta). Therefore, we suggest the following additional mitigation mechanisms.



\textbf{NVMe-oF Message Authentication.}
The NVMe-oF protocol could employ application-layer security to authenticate NVMe-oF messages including one-sided RDMA requests, even though source authentication for RDMA cannot be fully implemented at the application layer~\cite{srdma}. The reason is that one-sided RDMA requests are executed by RNICs without CPU involvement, making it impossible to verify the authenticity of these packets before executing them. For example, it is impossible to secure an RDMA read at the application layer as a remote client can access memory without CPU involvement using the DMA engine of the RNIC. 

Nonetheless, since NVMe-oF read and write requests always involve two-sided RDMA sends source authentication at the application layer can be used for these requests. We propose that the sender (i.e, an NVMe-oF client or an NVMe-oF target) includes a message authentication code (MAC) to each NVMe-oF message, thereby allowing the receiver to authenticate the sender and the data sent with a request (e.g., for write in-capsule). In this way, we can secure NVMe-oF requests and responses at the application layer. To additionally ensure data integrity of data sent over one-sided RDMA operations, we propose to authenticate this data with an additional MAC sent in a corresponding NVMe-oF message. 
For NVMe-oF writes, the MAC calculated over the data of the RDMA write should be part of the NVMe request. 
For NVMe-oF reads, the MAC calculated over the data of the RDMA read should be in the NVMe response.

To have such protection across different RDMA requests, the NVMe-oF application needs to invalidate memory registrations after completion of a one-sided RDMA request, preventing memory mutation caused by RDMA write injections from remote adversaries. After the memory is invalidated, the recipient of the data should verify the integrity of the data using a MAC from an NVMe-oF request or response. Otherwise, the attacker could modify the data after it was checked but before it was submitted to the local NVMe device. Note that the current implementation of SPDK chooses performance over security and does not invalidate memory registrations at all.
Overall, the NVMe-oF specification should introduce two MACs per an NVMe-oF message to authenticate data and NVMe-oF messages independently. If they use one MAC calculated over a concatenation of the message and the data, it may cause a partial execution of a falsified NVMe-oF write request as the NVMe-oF target needs to issue an RDMA read to fetch the data. 


\subsubsection{Guidelines for NVMe-oF Developers}
The SPDK library should register the target's memory only for local accesses and utilize invalidation of memory to significantly reduce the probability of unauthorized memory accesses. 
NVMe-oF developers could provide message authentication for RDMA connections using the aforementioned suggestions.
In this case, the in-band security should be extended to work with the RDMA connections. In addition, the in-band security should support challenging disconnect requests, allowing to verify the authenticity of RDMA disconnect requests. 
Developers could also randomize the RDMA identifiers with techniques proposed in other works~\cite{redmark}. Finally, 
developers should inform users of the NVMe-oF protocol about RDMA controllers and explain how to make use of them. 

\subsubsection{Guidelines for RDMA Vendors}
To prevent injection from local unprivileged users (as in \ta), 
the RDMA providers should introduce changes to the specification that add the source QPN to the packet format.
Furthermore, due to the shortcomings in IPsec over RoCE, RDMA should invest in a secure transport for InfiniBand-based protocols (e.g., sRDMA~\cite{srdma}).

The RDMA connection manager should process only messages arriving from other remote RDMA connection managers.
Finally, the RDMA  connection manager kernel module should be extended with a secure connection protocol that provides both secure RDMA channels and authentication of NVMe-oF (as discussed above). Currently, the RDMA  connection manager sends all data in plain text, thereby, forcing users of a future secure RDMA transport to implement their own secure connection establishment protocol.

\section{Related Work}
\label{sec:related}

Simpson et al.~\cite{Simpson} explore the security challenges introduced by RDMA networking to distributed storage systems. They analyze security gaps in RDMA techniques and their security implications for storage systems. However, their paper does not implement or test any of the suggested security gaps.

Rothenberger et al.~\cite{redmark} discuss vulnerabilities related to RDMA-enabled systems and implement several attacks on RDMA-enabled applications.  We extend their study in multiple dimensions. First, we relax their threat models for the injection attack and show how to perform an injection without administrative privileges on both RoCE and InfiniBand networks. Our injection tool is superior to ReDMArk in both performance and applicability (see Table~\ref{tab:comparison}). Second, we perform a security analysis of the RDMA connection manager, used by all RDMA systems in production, revealing vulnerabilities in its generation of connection keys. Third, we perform a thorough analysis of the NVMe-oF protocol in the scope of RDMA. 
Finally, we show an attack on the congestion mechanisms of RDMA, allowing to slow down a remote RNIC. 

Most of ReDMArk's attacks are not directly applicable to NVMe-oF, because NVMe-oF target applications can reject RDMA connections using IP filters or the in-band security extension (see Section~\ref{sec:inband}). Thus, ReDMArk's  connection exhaustion attacks and connection slowdown attacks by injecting traffic are not feasible. Overall, six out of seven  NVMe-oF attacks proposed in our work are not discussed in ReDMArk or not feasible with ReDMArk tools under the TLU model. Only ReDMArk's connection disconnection attack, which breaks connection by injecting invalid RDMA packets, is possible. Therefore, we discuss it in Section~\ref{sec:invalid}. 

Xing et al.~\cite{bedrock} develop a suite of defenses, called Bedrock, against attacks proposed in ReDMArk~\cite{redmark}. Bedrock leverages programmable data planes (using eBPF~\cite{ebpf} and P4~\cite{p4}) in modern network devices to build defense primitives for authentication, access control, and monitoring and logging in RDMA networks.  Their authentication tool can be employed to mitigate our injection attack, but it requires specialized programmable network controllers and switches, that are not always available. 
Bedrock cannot mitigate against our attack on the congestion control mechanisms as injected packets are not distinguishable from real congestion packets. Bedrock monitors traffic in switches, and thus cannot mitigate our local exhaustion attack and local connection key probing, as they do not generate network traffic. Nonetheless, Bedrock could be extended to detect local attacks by extending its eBPF framework to log local RDMA library calls.

An alternative technology for storage disaggregation is virtualization of remote NVMe storage via software-defined network accelerated processing~\cite{snap}. 
NVIDIA BlueField SmartNICs~\cite{bluefiled} enable hardware-accelerated virtualization of NVMe storage, where remote networked storage is emulated as a local NVMe SSD connected to the PCIe bus.  
As a result, the host operating system makes use of its standard NVMe-driver unaware that the NVMe access occurs over the network.
Their design avoids sharing RDMA drivers between the operating system and the SmartNIC, preventing the attacks under the \ta model and the ability to guess RDMA resources such as connection identifiers and keys. 

\section{Conclusion}
We show how we can bypass security mechanisms of NVMe-oF using vulnerabilities in RDMA protocols.
To perform attacks on NVMe-oF implementations we have designed several attack tools that can be used to 
attack other RDMA-enabled applications. Notably, we show how to spoof RDMA packets into victim connections without administrative permissions, even when IPsec over RoCE is enabled. 
Regarding the NVMe-oF  protocol, we show how an attacker without administrative permissions can write data to a remote NVMe device, bypassing existing security mechanisms of the NVMe-oF protocol and operating systems.
In addition, we show how we can falsify an NVMe data response, thereby forcing a victim client to observe the forged state of an NVMe device without actually modifying the state of the NVMe device.  
We anticipate that our work motivates security research on high-performance interconnects and systems utilizing them, leading to more secure high-performance networks and systems.

\section{Responsible Disclosure}
We have notified and responsibly disclosed the weaknesses related to RDMA and NVMe-oF to
Mellanox, Broadcom, Intel, and the NVMe working group prior to the submission
of this work.

\section{Availability of NeVerMore}
The source code of NeVerMore's attack tools is available upon request.

 
{\bibliographystyle{ACM-Reference-Format}
\bibliography{references}}


\end{document}